\documentclass[aps,pra,reprint,superscriptaddress,showpacs,longbibliography]{revtex4-1}
\usepackage[urlcolor=blue, hyperindex, colorlinks, bookmarks=true]{hyperref}
\usepackage{graphicx}
\usepackage{times}
\usepackage{amsmath}
\usepackage{braket}
\usepackage{xcolor}
\usepackage{tabularx}
\usepackage[symbol]{footmisc}

\usepackage{bm}
\usepackage{bbm}
\usepackage{natbib}
\usepackage{url}
\usepackage{textcase}
\usepackage{mhchem}

\definecolor{new}{rgb}{.38,.6,.38}
\definecolor{old}{rgb}{1,0,0}

\begin{document}

\title{Long-Range Microwave Mediated Interactions Between Electron Spins} 
\author{F.~Borjans}
\affiliation{Department of Physics, Princeton University, Princeton, New Jersey 08544, USA}
\author{X.~G.~Croot}
\affiliation{Department of Physics, Princeton University, Princeton, New Jersey 08544, USA}
\author{X.~Mi}
\altaffiliation{Present address: Google Inc., Santa Barbara, California 93117, USA}
\affiliation{Department of Physics, Princeton University, Princeton, New Jersey 08544, USA}
\author{M.~J.~Gullans}
\affiliation{Department of Physics, Princeton University, Princeton, New Jersey 08544, USA}
\author{J.~R.~Petta}
\email{petta@princeton.edu}
\affiliation{Department of Physics, Princeton University, Princeton, New Jersey 08544, USA}
\date{\today}

\begin{abstract}
{
Entangling gates for electron spins in semiconductor quantum dots are generally based on exchange, a short-ranged interaction that requires wavefunction overlap. Coherent spin-photon coupling raises the prospect of using photons as long-distance interconnects for spin qubits. Realizing a key milestone for spin-based quantum information processing, we demonstrate microwave-mediated spin-spin interactions between two electrons that are physically separated by more than 4 mm. Coherent spin-photon coupling is demonstrated for each individual spin using microwave transmission spectroscopy. An enhanced vacuum Rabi splitting is observed when both spins are tuned into resonance with the cavity, indicative of a coherent spin-spin interaction. Our results demonstrate that microwave-frequency photons can be used as a resource to generate long-range two-qubit gates between spatially separated spins.
}
\end{abstract}

\maketitle

Nonlocal qubit interactions are a hallmark of advanced quantum information technologies \cite{duan_long-distance_2001,majer_coupling_2007,sillanpaa_coherent_2007,monroe_scaling_2013,axline_-demand_2018}. The ability to transfer quantum states and generate entanglement over distances much larger than qubit length scales greatly increases connectivity and is an important step towards maximal parallelism and the implementation of two-qubit gates on arbitrary pairs of qubits \cite{preskill_john_reliable_1998}. Qubit coupling schemes based on cavity quantum electrodynamics \cite{majer_coupling_2007,wallraff_strong_2004,van_woerkom_microwave_2018} also offer the possibility of using high quality factor resonators as quantum memories \cite{sillanpaa_coherent_2007,pfaff_controlled_2017}. Furthermore, extending  qubit interactions beyond the nearest neighbor is particularly beneficial for spin-based quantum computing architectures limited by short-range exchange interactions \cite{petta_coherent_2005}.

Silicon spin qubits can be fabricated in dense arrays and the device technology has matured rapidly in the past several years, with single qubit gate fidelities that rival superconducting qubits \cite{yoneda_quantum-dot_2018}, demonstrations of exchange-based two qubit gates \cite{veldhorst_two-qubit_2015,zajac_resonantly_2018,watson_programmable_2018}, and promising initial steps towards quantum information transfer along qubit arrays \cite{fujita_coherent_2017,mills_shuttling_2019}. Long-range coherent coupling between separated semiconductor spin qubits is an outstanding challenge. Efforts towards nonlocal spin-spin coupling have been limited to interactions between individual spins and photons in a microwave cavity \cite{viennot_coherent_2015,mi_coherent_2018,samkharadze_strong_2018,landig_coherent_2018}.

One of the key challenges associated with achieving long distance spin-spin coupling stems from the lack of independent control of the spin Zeeman energies. In order to coherently couple spin qubits via cavity photons \cite{warren_long-distance_2019,benito_optimized_2019}, it is necessary to bring the qubits into resonance with each other (virtual coupling), or into resonance with each other $and$ the microwave cavity (resonant coupling). Unlike charge qubits, whose energy splittings can easily be electrically tuned \cite{van_woerkom_microwave_2018}, spin qubits confined in quantum dots have limited electrical tunability. Their energies are set by a total magnetic field $\vec B^{\rm tot}$, which results from the vector addition of a global external field $\vec B^{\rm ext}$ and stray fields from micromagnets $\vec B^{\rm M}$ deposited on-chip to facilitate spin-photon coupling \cite{mi_coherent_2018,benito_input-output_2017}. Small variations in micromagnet fabrication at each qubit lead to variations in $\vec B^{\rm tot}$, making resonant coupling of spins challenging. While spin-charge hybridization permits a small degree of frequency tunability, generally it is insufficient to simultaneously tune spin qubits into resonance with each other and the cavity. Overcoming these challenges, we use a microwave frequency photon to coherently couple two electron spins that are separated by more than 4 mm, opening the door to a Si spin qubit architecture with all-to-all connectivity. Deliberately engineered asymmetries in micromagnet design, in conjunction with three-axis control of the external magnetic field, allow us to tune separated spin qubits into resonance with each other and the cavity.

\begin{figure}[htb]
\centering
\includegraphics[width=1.0\columnwidth]{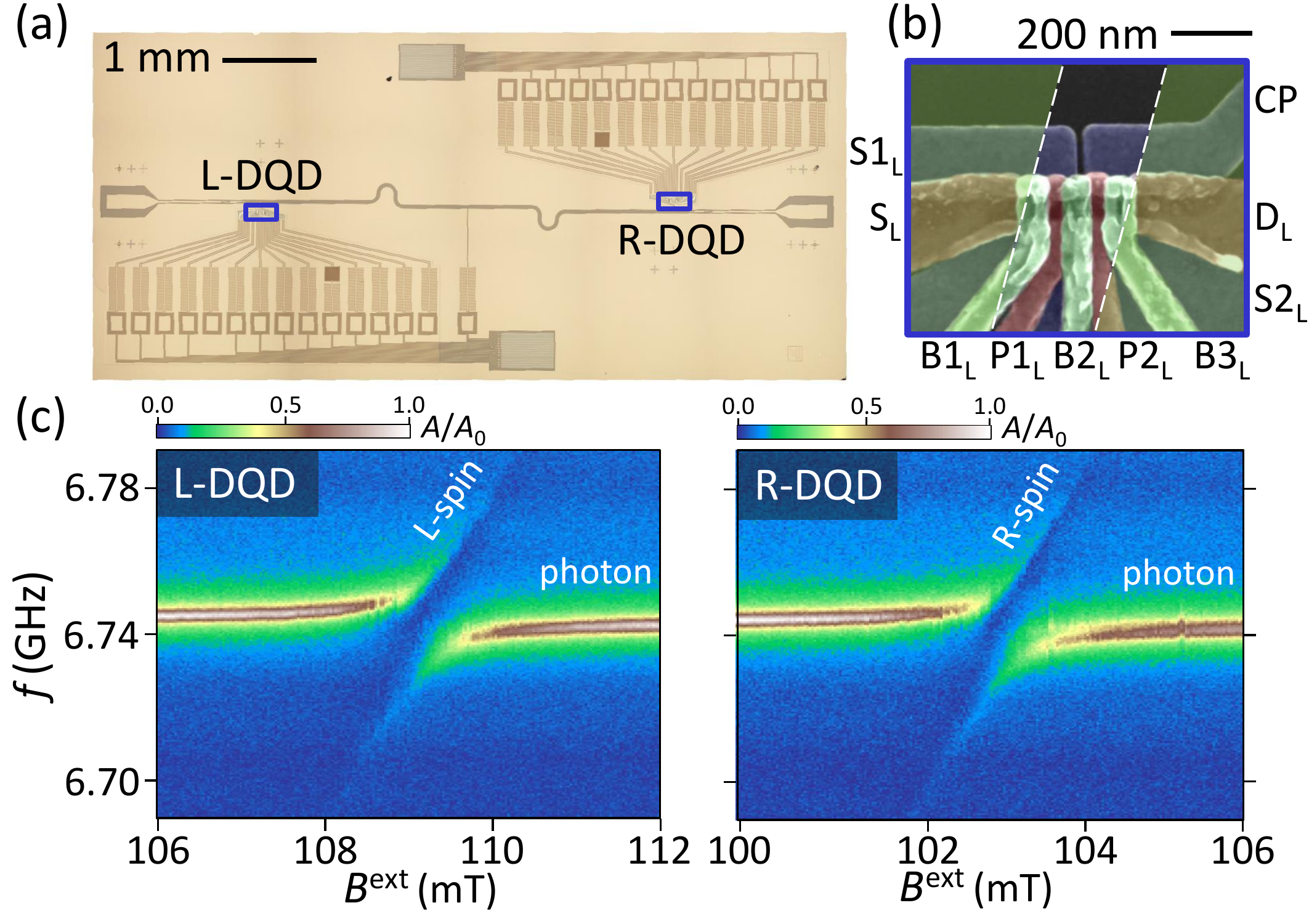}
\caption{Cavity-coupler for spins. (a) Optical micrograph of the superconducting cavity containing two single electron DQDs. The electron spin in each DQD is coupled to the cavity through a combination of electric-dipole and artificial spin-orbit interactions. (b) False-color scanning electron microscope image of the L-DQD. A double well potential is formed beneath plunger gates $\rm P1_L$ and $\rm P2_L$, and the barrier gate $\rm B2_L$ is used to adjust interdot tunnel coupling. Spin-orbit coupling is induced by a Co micromagnet (dashed lines). (c) Cavity transmission $A/A_0$ plotted as a function of the external magnetic field $B^{\rm ext}$ for the L-DQD and R-DQD. Vacuum Rabi splitting, a hallmark of strong coupling, is observed for each spin.}
\label{fig:1}
\end{figure}

The device consists of two double quantum dots (DQDs), denoted L-DQD and R-DQD, that are fabricated on a Si/SiGe heterostructure and positioned at the antinodes of a half-wavelength Nb superconducting cavity [Fig.\ \ref{fig:1}(a)] with a center frequency $f_{c} = 6.745\,\rm GHz$ and decay rate $\kappa/2\pi = 1.98\,\rm MHz$. Device 1 is fabricated on a nat-Si quantum well, while Device 2 utilizes an enriched \ce{^{28}Si} quantum well with an 800 ppm residual concentration of \ce{^{29}Si}. A single electron is isolated in each DQD, which is defined by a tri-layer overlapping aluminum gate stack \cite{zajac_scalable_2016} operated in accumulation mode [Fig.\ \ref{fig:1}(b)]. The electron interacts with the electric field of the cavity through the electric dipole interaction. Our device design uses a split-gate cavity coupler [labelled CP in Fig.\ \ref{fig:1}(b)] that is galvanically connected to the center pin of the superconducting cavity \cite{petta_prep_2019}. The split-gate coupler relaxes the constraint that plunger gates be held at the same potential \cite{mi_coherent_2018} and allows simultaneous tuning of both DQDs to the one electron regime, which is necessary for a demonstration of spin-spin coupling. Measurements take place in a dilution refrigerator and an in-plane external magnetic field $B^{\rm ext}$ is used to tune the spin transition into resonance with a cavity photon. 

The small magnetic field generated by the vacuum fluctuations of the cavity, combined with the weak magnetic moment of the electron $\mu_B \approx 58\,\rm \mu eV/T$, leads to an intrinsic spin-photon coupling rate $g_s/2\pi \approx 10 – 100\,\rm Hz$, which is much too slow to be useful as a quantum technology \cite{schuster_high-cooperativity_2010,kubo_strong_2010}. In our device architecture a large electric dipole coupling rate $g_c/2\pi \approx 40\,\rm MHz$ is combined with an artificial spin-orbit interaction generated by a micromagnet to achieve spin-photon coupling \cite{viennot_coherent_2015,cottet_spin_2010,benito_input-output_2017}. The resulting spin-cavity coupling rate is given by $g_s\propto g\mu_B\Delta B_x$, where $g \approx 2$ is the electronic g-factor and $\mu_B$ is the Bohr magneton and $\Delta B_x$ is the amplitude of the effective oscillating transverse field generated by the electron’s motion in the cavity field \cite{mi_coherent_2018,samkharadze_strong_2018,benito_input-output_2017}. $\Delta B_x$ is maximal at the interdot charge transition, where the level detuning $\epsilon = 0$, and $g_c/2\pi$ is largest. Large gradient fields have been used as a resource to enable high fidelity single spin manipulation \cite{pioro-ladriere_electrically_2008,yoneda_robust_2015} and recent demonstrations of single spin-photon coupling with $g_s/2\pi \approx 10\,\rm MHz$ \cite{mi_coherent_2018,samkharadze_strong_2018}.

We first demonstrate strong coupling of a spin trapped in each DQD to a cavity photon. To probe spin-photon coupling, the cavity transmission $A/A_0$ is plotted as a function of cavity probe frequency $f$ and $B^{\rm ext}$ in Fig.\ \ref{fig:1}(c). The Zeeman splitting increases with the total magnetic field $\vec{B}^{\rm tot} = \vec{B}^{\rm ext}+\vec{B}^{\rm M}$. Field tuning allows us to bring each spin into resonance with the cavity photon of energy $h f_c$, where $h$ is Planck’s constant. We observe coherent coupling between the spin trapped in the L-DQD (L-spin) and the cavity photon, as evidenced by the vacuum Rabi splitting, when $B^{\rm ext} = 109.1\,\rm mT$. Strong spin-photon coupling is achieved with $g_{s,L}/2\pi = 10.7\pm 0.1\,\rm MHz$ exceeding the cavity decay rate $\kappa/2\pi = 1.98\,\rm MHz$ and spin decoherence rate $\gamma_{s,L}/2\pi = 4.7\,\rm MHz$ (Appendix \ref{app:B}). Likewise, we observe strong spin-photon coupling for the R-spin at $B^{\rm ext} = 103.1\,\rm mT$ with $g_{s,R}/2\pi = 12.0\pm 0.2\,\rm MHz$ exceeding $\gamma_{s,R}/2\pi =  5.3\,\rm MHz$ and $\kappa/2\pi$. The $6\,\rm mT$ difference in $B^{\rm ext}$ is equivalent to a $268\,\rm MHz$ difference in the spin resonance frequencies after accounting for the finite susceptibility of the micromagnet (Appendix \ref{app:B}), and is largely due to inaccuracies in micromagnet placement. This substantial frequency detuning between the spins precludes the observation of resonant spin-spin coupling \cite{mi_coherent_2018}. 
	
\begin{figure}[t!]
\includegraphics[width=1.0\columnwidth]{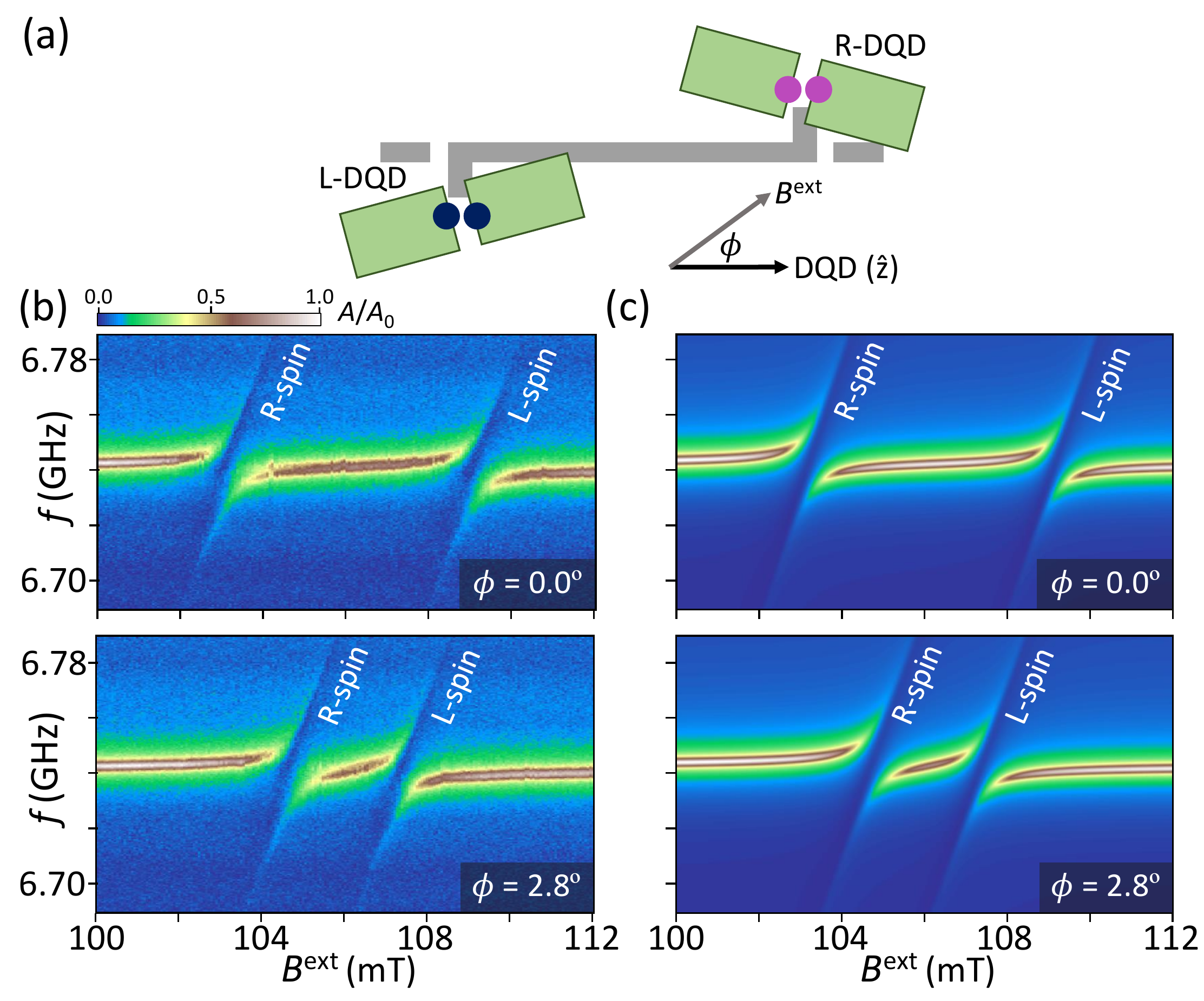}
\caption{Tuning towards spin-spin resonance. (a) The micromagnets are purposely fabricated at an angle relative to the DQD z-axis to allow for additional control of the spin resonance conditions. By rotating $B^{\rm ext}$ relative to the z-axis, and tuning the magnitude of $B^{\rm ext}$, the energy splitting of both spins can be brought into resonance with the cavity mode. (b) $A/A_0$ as a function of $B^{\rm ext}$ and $f$. For a field angle $\phi=0^\circ$ the R-spin and L-spin resonance conditions are separated by $B^{\rm ext} = 6\,\rm mT$. Increasing the field angle to $\phi = 2.8^\circ$ dramatically reduces the field separation. (c) Theoretical model for $A/A_0$ using a master equation simulation.}
\label{fig:2}
\end{figure}	

\begin{figure*}[t!]
\includegraphics[width=1.5\columnwidth]{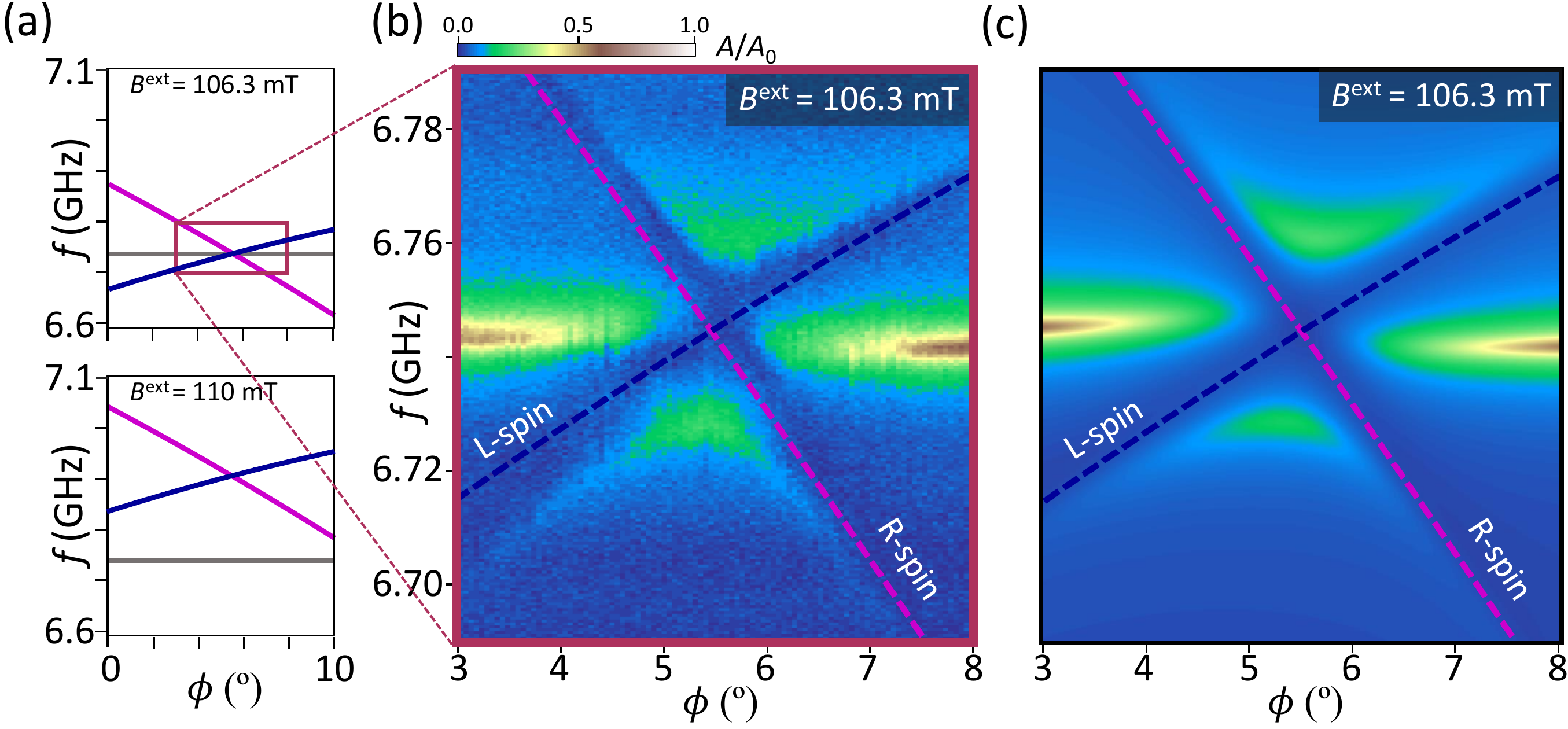}
\caption{(a) Expected spin resonance frequencies as a function of $\phi$ for $B^{\rm ext} = 106.3\,\rm mT$ (top panel) and $B^{\rm ext} = 110\,\rm mT$ (bottom panel). $B^{\rm ext}$ allows control over both spin frequencies with respect to the cavity, while $\phi$ allows control over the spin frequencies with respect to each other. The frequency of the left (right) spin is plotted in blue (purple). (b) $A/A_0$ as a function of $f$ and $\phi$ demonstrates simultaneous tuning of both spins into resonance with the cavity at $\phi = 5.6^\circ$ and $B^{\rm ext} = 106.3\,\rm mT$. Dashed lines indicate left and right spin transition frequencies. (c) Theoretical prediction for $A/A_0$.}
\label{fig:3}
\end{figure*}
	
To compensate for local differences in the magnetic field, the device was purposely fabricated with the long axis of the Co micromagnets tilted $\pm 15$ degrees relative to the interdot axis of the DQDs [Fig.\ \ref{fig:2}(a)]. Because of this intentional asymmetry, adjusting the angle $\phi$ of the in-plane magnetic field relative to the DQD axis provides an additional degree of freedom for simultaneous tuning of both spins into resonance with the cavity. Qualitatively, the high permeability Co micromagnet concentrates the magnetic field lines, leading to a maximum total field when $\vec{B}^{\rm ext}$ is aligned with the long axis of the micromagnet. A similar approach was adopted to remotely couple two ensembles of nitrogen vacancy centers in diamond, where the crystal axes of the two diamond samples were rotated relative to one another \cite{astner_coherent_2017}. We show now that this approach is well-suited for coupling two single spins in silicon via a cavity mode.

Figure \ref{fig:2}(b) shows $A/A_0$ as a function of $f$ and $B^{\rm ext}$ with $\phi = 0^\circ$. The R-spin (L-spin) is in resonance with the cavity mode at $B^{\rm ext} \approx 103.1\,\rm mT$ ($B^{\rm ext} \approx 109.1\,\rm mT$). With $\phi = 2.8^\circ$, the R-spin resonance condition shifts up to $B^{\rm ext} \approx 104.7\,\rm mT$ and the L-spin resonance condition shifts down to $B^{\rm ext} \approx 107.4\,\rm mT$. To compare the data with theory we model the system with an effective Jaynes-Cummings Hamiltonian \cite{mi_coherent_2018,benito_input-output_2017}
\begin{equation}
\label{eq:1}
H=\sum\limits_j \frac{\hbar \omega_j}{2}\sigma_j^z+\hbar\omega_c a^\dagger a + \hbar g_{s,j}\left(\sigma_j^+ a + \sigma_j^- a^\dagger\right),
\end{equation}
where $j=L,R$, and $\hbar\omega_j$ is the energy level splitting of the effective $j$-spin, $\sigma_j^\mu$ are the Pauli operators in the $j$-spin Hilbert space and $\omega_c/2\pi$ is the cavity frequency. Decoherence in our system is incorporated by a Lindblad master equation (Appendix \ref{app:A}), using the measured $\gamma_{s,j}$ and $\kappa$. We plot the theoretical predictions for the cavity transmission in Fig. \ref{fig:2}(c) and observe strong agreement with the experimental data.

We now demonstrate control over the difference between the spin resonance frequencies by fixing the external magnetic field magnitude $B^{\rm ext}$ and varying the angle of $B^{\rm ext}$ in the plane of the sample. The expected spin resonance frequencies are plotted in Fig.\ \ref{fig:3}(a) as a function of $\phi$ with $B^{\rm ext} = 106.3\,\rm mT$ (upper panel) and $B^{\rm ext} = 110\,\rm mT$ (lower panel), confirming that these two control parameters will bring the two spins into resonance with the cavity and each other. Based on microwave spectroscopy measurements of the spins, we expect resonance to occur around $\phi = 6^\circ$ and $B^{\rm ext} = 106.3\,\rm mT$. Figure \ref{fig:3}(b) maps out the field angle dependence of the spin resonance frequencies over a range $\phi= 3–8^\circ$ with $B^{\rm ext} = 106.3\,\rm mT$. The resonance frequency of the R-spin monotonically moves to lower $f$ as $\phi$ is increased, while the L-spin shows the opposite dependence. With $\phi= 5.6^\circ$ both spins are tuned into resonance with the cavity. These results are well captured by the theoretical prediction of transmission through the cavity shown in Fig. \ref{fig:3}(c).

\begin{figure*}[t!]
\includegraphics[width=2.0\columnwidth]{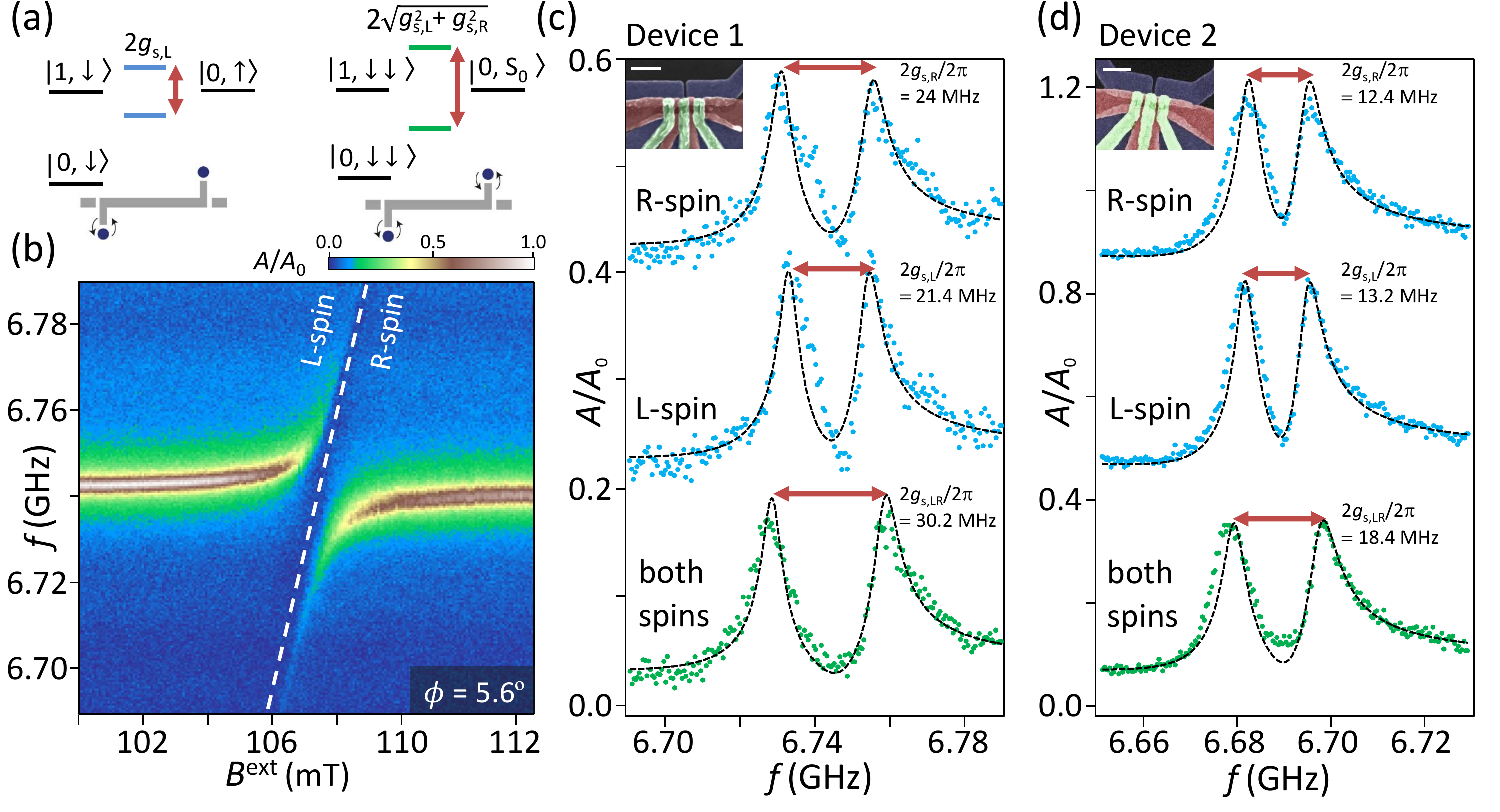}
\caption{Cavity-mediated spin-spin coupling. (a) Tuning the L-spin into resonance with the cavity results in vacuum Rabi splitting with magnitude $2g_{s,L}$. A vacuum Rabi splitting of magnitude $2g_{s,LR}=2\sqrt{g_{s,L}^2+g_{s,R}^2}$ is expected when both spins are tuned into resonance with the cavity. (b) $A/A_0$ as a function of $f$ and $B^{\rm ext}$ with $\phi = 5.6^\circ$ indicates an enhanced vacuum Rabi splitting when the L-spin and R-spin are tuned into resonance with the cavity. (c) $A/A_0$ as a function of $f$ for the R-spin in resonance (upper curve), L-spin in resonance (middle curve), and both spins in resonance (bottom curve). The enhancement of the vacuum Rabi splitting with both spins on resonance with the cavity is indicative of spin-spin coupling via the cavity mode. (d) Cavity-assisted spin-spin coupling is also observed in a second device with a different gate pattern. Dashed lines in (c,d) are fits to a master equation simulation. Insets in (c,d) are scanning electron microscope images of the devices, with $200\,\rm nm$ scale bars.}
\label{fig:4}
\end{figure*}

The spectrum of the Jaynes-Cummings model for a single spin and a single photon in the cavity is shown in Fig.~\ref{fig:4}(a). With the L-spin tuned into resonance with the cavity, the spin and cavity photon hybridize leading to a vacuum Rabi splitting of magnitude $2g_{s,L}$ in the cavity transmission \cite{wallraff_strong_2004}. In contrast, when both spins are simultaneously tuned into resonance with the cavity, the excited state spectrum of the Jaynes-Cummings model splits into a sub-radiant state and two bright states \cite{majer_coupling_2007}. For N identical spins the Jaynes-Cummings model predicts a $\sqrt{N}$ enhancement of the coupling rate \cite{wallraff_strong_2004}. In our device geometry, the sub-radiant state is the spin-triplet $\ket{0,T_0}=1/\sqrt{2}\left(\ket{0,\uparrow ,\downarrow}+\ket{0,\downarrow , \uparrow}\right)$ because the DQDs are located at opposite ends of the cavity where the electric fields are $180^\circ$ out of phase. Here, the spin states of the L/R-spin are quantized along their local total magnetic field axis. The bright states are hybridizations between the singlet state $\ket{0,S_0}=1/\sqrt{2}\left(\ket{0,\uparrow ,\downarrow}-\ket{0,\downarrow , \uparrow}\right)$ and the state with a single photon $\ket{1,\downarrow , \downarrow}$ that are separated in energy by twice the collectively enhanced vacuum Rabi coupling, $2g_{s,LR}=2\sqrt{g_{s,L}^2+g_{s,R}^2}$.  We now search for evidence of cavity-mediated single spin coupling.

Figure \ref{fig:4}(b) shows $A/A_0$ as a function of $f$ and $B^{\rm ext}$ with $\phi = 5.6^\circ$, where both spins are in resonance with the cavity. As $B^{\rm ext}$ is increased, both spins are simultaneously tuned into resonance with the cavity at $B^{\rm ext} = 106.3\,\rm mT$ and we observe an enhancement in the vacuum Rabi splitting relative to the data sets shown in Fig.\ \ref{fig:2}(b). The vacuum Rabi splitting is quantitatively analyzed for Device 1 in Fig.\ \ref{fig:4}(c) and Device 2 in Fig.\ \ref{fig:4}(d). These devices have slight differences in gate geometry, and the micromagnets in Device 2 are canted at angles of $\pm 10$ degrees. In Device 1, we extract $g_{s,L/R}$ by measuring $A/A_0$ at $\phi = 5.6^\circ$ and detuning the right/left spin using $\epsilon$. In Device 2, we extract the $g_{s,L/R}$ by measuring $A/A_0$ at an off-resonant angle, at the now separated resonant fields. For Device 1, we observe a vacuum Rabi splitting of $2g_{s,L}/2\pi$ = $21.4\pm 0.2\,\rm MHz$ when L-spin is in resonance with the cavity, $2g_{s,R}/2\pi = 24.0\pm 0.4\,\rm MHz$ when R-spin is in resonance with the cavity, and $2g_{s,LR}/2\pi = 30.2\pm 0.2\,\rm MHz$ when both spins are in resonance with the cavity. Device 2 shows similar behavior and again exhibits an enhanced vacuum Rabi splitting $2g_{s,LR}/2\pi = 18.4\pm 0.4\,\rm MHz$ when both spins are tuned into resonance with the cavity, compared to the individual splittings $2g_{s,L}/2\pi = 13.2\pm 0.2\,\rm MHz$ and $2g_{s,R}/2\pi = 12.4\pm 0.2\,\rm MHz$. Combined, these two data sets give strong evidence for microwave assisted spin-spin interactions across a $4\,\rm mm$ length scale that is many orders of magnitude larger than what can be achieved using direct wavefunction overlap. Moreover, these measurements show that both field angle and DQD level detuning can be used to modulate the strength of the spin-spin interactions. 

The data in Fig.\ \ref{fig:4}(c) are fit using a master equation description of the spin-cavity system.  We independently measure the Zeeman splittings $\hbar\omega_{L/R}$ and linewidths $\gamma_{s,L/R}$ using microwave spectroscopy (Appendix \ref{app:B}). The cavity linewidth $\kappa$, as well as a complex Fano factor $q$, are obtained by fitting the bare cavity response with the spins detuned from resonance (Appendix \ref{app:B}).  The spin-photon coupling rates $g_{s,L/R}$ for each device are obtained by fitting the data with the spins individually tuned into resonance with the cavity as shown in Figs.\ \ref{fig:4}(c),(d). From the Jaynes-Cummings model we expect the bright states to split with the enhanced collective coupling rate $2g_{s,LR}=2\sqrt{g_{s,L}^2+g_{s,R}^2}$. The extracted splittings agree with the theoretical predictions of $2g_{s,LR}/2\pi = 32.1\,\rm MHz$ for Device 1 and $2g_{s,LR}/2\pi = 18.1\,\rm MHz$ for Device 2 to within 6\%.

The observation of enhanced vacuum Rabi splitting when the separated spins are simultaneously on resonance with the cavity is evidence of long-range spin-spin coupling. The nonlocal interaction of two spins marks an important milestone for all-to-all qubit connectivity and scalability in silicon-based quantum circuits. In the near term, this demonstration paves the way for the implementation of modular qubit architectures in silicon, wherein nearest-neighbor coupled registers of spin qubits \cite{zajac_scalable_2016} can be interfaced with other sparsely distributed registers via microwave photons. With further improvements in cavity quality factors and spin-photon coupling rates, time-domain demonstrations of cavity-assisted spin-spin coupling should be within experimental reach.

\begin{acknowledgements}
We thank Lisa Edge, Joe Kerckhoff, Thaddeus Ladd, and Emily Pritchett of HRL Laboratories, LLC for providing the \ce{^{28}Si} heterostructure used in these experiments, device simulation support, and technical comments on the manuscript. We acknowledge helpful discussions with Monica Benito and Guido Burkard.
\end{acknowledgements}

\appendix
\section{Master Equation Theory}
\label{app:A}
Since the transverse magnetic field $\Delta B_x\approx 30\,\rm mT$ is large compared to the charge-cavity coupling rate $g_c$, we can use the effective Hamiltonian described in Equation \ref{eq:1}. We introduce dephasing in our system by coupling the spin-cavity system to external baths. Integrating out these baths in a Born-Markov approximation leads to a Lindblad master equation for the spin-cavity dynamics
\begin{equation}
\frac{d\rho}{dt}=-\frac{i}{\hbar}[H,\rho]+\sum\limits_j\left(\gamma_{s,j}^d\mathcal{D}\left[\sigma_j^z\right]+\gamma_{s,j}^r\mathcal{D}\left[\sigma_j^-\right]\right)\rho + \kappa \mathcal{D}\left[\alpha\right]\rho
\end{equation}
where $\mathcal{D}[A]=A\rho A^\dagger - \frac{1}{2}(A^\dagger A\rho + \rho A^\dagger A)$ is the Lindblad super-operator, $\gamma_{s,j}^d$ is the spin-dephasing rate, $\gamma_{s,j}^r$ is the relaxation rate of the spins, and $\kappa$ is the total cavity decay rate. The spin linewidth (FWHM) is given by $2\gamma_{s,j}=2\gamma_{s,j}^d+\gamma_{s,j}^r$. We decompose $\kappa =\kappa_1+\kappa_2+\kappa_{in}$ into the contributions from the decay rates into the two transmission lines $\kappa_{1,2}$ and an intrinsic cavity decay rate $\kappa_{in}$.

In this experiment, DQDs are probed via the cavity transmission. The transmission can be computed from the master equation evolution using input-output theory \cite{meystre_elements_2007}, which relates the output field operator $a_{n,out}$ of each transmission line to the input field operator $a_{n,in}$ through the relation
\begin{equation}
a_{n,out}(t)= \sqrt{\kappa_n} a(t)-a_{n,in} (t)
\end{equation}
where $a(t)$ is found from the operator evolution under the Lindblad master equation with the addition of a drive term $H_{drive}=\sum_n\sqrt{\kappa_n}\left(a^\dagger a_{n,in}+h.c.\right)$.  In the limit of weak driving $\sqrt{\kappa_n}\braket{a_{n,in}} \ll \kappa$ (here operator expectation values of the input-output fields are taken with respect to the density matrix of the bath, which is treated as unentangled with the system in the Born-Markov approximation), we can neglect the population of the excited states of the Jaynes-Cummings model with more than one excitation, leaving the reduced spin-photon Hilbert space
\begin{equation}
\ket{0,\downarrow,\downarrow},\quad \ket{0,\uparrow,\downarrow},\quad \ket{0,\downarrow,\uparrow},\quad \ket{1,\downarrow,\downarrow}
\end{equation}
where the first quantum number labels the number of photons in the cavity and the left/right spin labels refer to the L/R-spin direction quantized along the axis of the local magnetic field. When $g_{s,j} \mathrm{Tr}[\rho a] \ll \gamma_{s,j}$, we can further neglect populations outside of the ground state $\ket{0,\downarrow,\downarrow}$ and only need to account for off-diagonal coherences in the master equation evolution. In this limit, the complex transmission amplitude takes the approximate form 
\begin{equation}
A=\frac{\braket{a_{2,out}}}{\braket{a_{1,in}}}=\frac{\sqrt{\kappa_2}\mathrm{Tr}[\rho a]}{\braket{a_{1,in}}}=\frac{-i(\sqrt{\kappa_1\kappa_2}+\Delta_0/q)}{-\Delta_0-i\frac{\kappa}{2}-i\sum_j\frac{g_{s,j}^2}{\gamma_{s,j}+i\delta_j}}
\label{eq:inout}
\end{equation}
where $\Delta_0=\omega_d-\omega_c$ is the detuning of the cavity drive $\omega_d$ from the cavity frequency, $\delta_j=\omega_j-\omega_d$ is the detuning of the $j$-spin from the drive frequency, and we have introduced a complex parameter $q$ to account for Fano interference effects in the bare cavity transmission \cite{fano_effects_1961}.  For simplicity, we neglect all higher order corrections in $g_{s,j}/q$ to the cavity transmission.

\section{Theory Fits}
\label{app:B}
From the spin-photon transitions in Fig.\ \ref{fig:1}(c), we fit the qubit frequency and extract the finite magnetic susceptibility $\chi$ of the micromagnet, finding $\chi\approx 0.6$ at $\phi=0^\circ$ over small ranges of magnetic field, in agreement with previous work \cite{mi_coherent_2018}. Substituting the extracted qubit frequency $\omega_j$ into Eq. \ref{eq:inout}, we generate the theory plots in Fig. \ref{fig:2}(c).

To generate the fit in Fig.\ \ref{fig:3}(c), the angular dependence of the resonance frequencies of the two DQDs is extracted using microwave spectroscopy in the dispersive regime. We can then shift the dependence in frequency based on the external magnetic field and the known magnetic
susceptibility to overlay with the data. 

The fits to the line cuts shown in Figs.\ \ref{fig:4}(c),(d) are modeled with Equation \ref{eq:inout}, using the input parameters in Table \ref{tab:pars}. We leave $g_s$ as a free parameter in nonlinear regression fits and fix the remaining parameters extracted from separate measurements. The errors on $g_s$ given in the main text are determined from these fits. We assume the gradient field between the two dots is $\Delta B_x=30\,\rm mT$ based on previous experiments \cite{mi_coherent_2018}. We define $\pm\theta$ as the rotation angle of the micromagnet long axis relative to the interdot axis of the left (right) DQDs, and $t_{c,L/R}$ is the interdot tunnel coupling of the left (right) DQDs. To fit the data in Figs.\ \ref{fig:4}(c),(d) with both spins on resonance, we use an effective model where we extract the splitting of the two bright states using a single collective coupling $g_{s,LR}$ and a single effective decay rate $\gamma_{s,LR}$. These fits are in good agreement with the predicted curves (not shown) for the two-spin system based on equation \ref{eq:inout} with the parameters from
Table \ref{tab:pars}.

\begin{table}[b]
\caption{\label{tab:pars}Physical parameters of the two devices.}
\begin{ruledtabular}
 \begin{tabular}{ccc} 
 Parameter & Device 1 & Device 2 \\ 
 \hline
 $g_{s,L}/2\pi$ & $10.7\,\rm MHz$ & $6.6\,\rm MHz$ \\ 
 $g_{s,R}/2\pi$ & $12.0\,\rm MHz$ & $6.2\,\rm MHz$ \\
 $g_{s,LR}/2\pi$ & $15.1\,\rm MHz$ & $9.2\,\rm MHz$ \\
 $\gamma_{s,L}/2\pi$ & $4.7\,\rm MHz$ & $3.0\,\rm MHz$ \\
 $\gamma_{s,R}/2\pi$ & $5.3\,\rm MHz$ & $3.3\,\rm MHz$ \\
 $\gamma_{s,LR}/2\pi$ & $5.0\,\rm MHz$ & $4.0\,\rm MHz$ \\
 $2t_{c,L}/h$ & $8.8\,\rm GHz$ & $9.1\,\rm GHz$ \\
 $2t_{c,R}/h$ & $8.6\,\rm GHz$ & $9.2\,\rm GHz$ \\
 $\kappa/2\pi$ & $2.0\,\rm MHz$ & $3.46\,\rm MHz$ \\
 $q$ & $33.3e^{-0.3i\pi}$ & $20e^{-0.25i\pi}$ \\
 $\Delta B_x$ & $30\,\rm mT$ & $30\,\rm mT$ \\
 $\theta$ & $15^\circ$ & $10^\circ$ \\
 $\chi$ & $0.6$ & $0.6$ \\
\end{tabular}
\end{ruledtabular}
\end{table}

\bibliography{References_Borjans_PRX_2019_v2}

\end{document}